\documentclass[prl,preprint,superscriptaddress]{revtex4}
\usepackage[normalem]{ulem} 
\usepackage{bm}
\usepackage[labelfont=large,position=top]{subfig}
\usepackage{amsfonts}
\usepackage{amsmath}
\usepackage{amssymb}
\usepackage{mathrsfs}
\usepackage{ot1patch}
\usepackage{amsmath}

\usepackage{graphicx}
\usepackage{dcolumn}
\usepackage{bm}
\usepackage[thinspace,squaren,pstricks,italian,derivedinbase,derived]{SIunits}
\usepackage[colorlinks=true,citecolor=blue,linkcolor=blue,urlcolor=blue]{hyperref}

\begin{document}

\title{Coherent transfer of spin angular momentum by evanescent spin waves within antiferromagnetic NiO}  

\author{Maciej D\c{a}browski}
\email{M.K.Dabrowski@exeter.ac.uk}
\affiliation{Department of Physics and Astronomy, University of Exeter, Stocker Road, Exeter, Devon EX4 4QL, United Kingdom}
\author{Takafumi Nakano}
\affiliation{Department of Physics and Astronomy, University of Exeter, Stocker Road, Exeter, Devon EX4 4QL, United Kingdom}
\affiliation{Spintronics Research Center, National  Institute of Advanced Industrial Science and Technology (AIST), Tsukuba, 305-8568, Japan}
\author{David M. Burn}
\affiliation{Magnetic Spectroscopy Group, Diamond Light Source, Didcot, OX11 0DE, United Kingdom}
\author{Andreas Frisk}
\affiliation{Magnetic Spectroscopy Group, Diamond Light Source, Didcot, OX11 0DE, United Kingdom}
\author{David G. Newman}
\affiliation{Department of Physics and Astronomy, University of Exeter, Stocker Road, Exeter, Devon EX4 4QL, United Kingdom}
\author{Christoph Klewe}
\affiliation{Advanced Light Source, Lawrence Berkeley National Laboratory, Berkeley, California 94720, USA}
\author{Qian Li}
\affiliation{Department of Physics, University of California at Berkeley, California 94720, USA}
\author{Mengmeng Yang}
\affiliation{Department of Physics, University of California at Berkeley, California 94720, USA}
\author{Padraic Shafer}
\affiliation{Advanced Light Source, Lawrence Berkeley National Laboratory, Berkeley, California 94720, USA}
\author{Elke Arenholz}
\affiliation{Advanced Light Source, Lawrence Berkeley National Laboratory, Berkeley, California 94720, USA}
\author{Thorsten Hesjedal}
\affiliation{Department of Physics, Clarendon Laboratory, University of Oxford, OX1 3PU, United Kingdom}
\author{Gerrit van der Laan}
\affiliation{Magnetic Spectroscopy Group, Diamond Light Source, Didcot, OX11 0DE, United Kingdom}
\author{Zi Q. Qiu}
\affiliation{Department of Physics, University of California at Berkeley, California 94720, USA}
\author{Robert J. Hicken}
\email{r.j.hicken@exeter.ac.uk}
\affiliation{Department of Physics and Astronomy, University of Exeter, Stocker Road, Exeter, Devon EX4 4QL, United Kingdom}

\begin{abstract}
Insulating antiferromagnets are efficient and robust conductors of spin current. To realise the full potential of these materials within spintronics, the outstanding challenges are to demonstrate scalability down to nanometric lengthscales and the transmission of coherent spin currents. Here, we report the coherent transfer of spin angular momentum by excitation of evanescent spin waves of GHz frequency within antiferromagnetic NiO at room temperature. Using element-specific and phase-resolved x-ray ferromagnetic resonance, we probe the injection and transmission of \textit{ac} spin current, and demonstrate that insertion of a few nanometre thick epitaxial NiO(001) layer between a ferromagnet and non-magnet can even enhance the flow of spin current. Our results pave the way towards coherent control of the phase and amplitude of spin currents at the nanoscale, and enable the realization of spin-logic devices and spin current amplifiers that operate at GHz and THz frequencies.  
\end{abstract}

%

\maketitle

One of the most fascinating developments in spintronics has been the use of insulating antiferromagnets (AFMs) as media for the transport of spin angular momentum by spin waves \cite{Lebrun2018,Gomonay2018,Hou2019,Jungwirth2016,Qiu2018}. While robustness against magnetic perturbations and functionality in the terahertz (THz) frequency range \cite{Satoh2010,Kanda2011} are common to most AFMs, exceptionally long-distance spin transport \cite{Lebrun2018} and current induced switching of the N\'eel vector by antidamping torques \cite{Chen2018,Gray2019,Baldrati2019} has been experimentally proven only for insulating phases so far. Yet, experimental studies have been limited to transport of time-averaged \textit{dc} spin currents, providing no insight into the underlying dynamic processes within the antiferromagnet \cite{Li2019}. In particular, control of coherent spin currents, where the phase of a spin wave provides an additional degree of freedom with which to encode information \cite{Chumak2015}, remains to be demonstrated. The coherence of spin waves is expected to be especially important when downscaling magnonic media \cite{Yu2016}. It has been theoretically predicted by Khymyn \emph{et al.} \cite{Khymyn2016} that in the case of a few nanometre thick AFM with \emph{biaxial} anisotropy, coherent excitation of evanescent spin waves can transfer angular momentum from the AFM lattice to the spin subsystem, resulting in the amplification of the transmitted spin current. Although enhanced \textit{dc} spin current transmission through thin ($\leq$\,6\,nm) NiO films has been detected by inverse spin Hall effect (ISHE) measurements \cite{Wang2014,Hahn2014,Lin2016}, the microscopic mechanism is still under debate, since alternative explanations based upon diffusion of thermal magnons can also reproduce the measured dependence of the \textit{dc} spin current on the NiO thickness \cite{Rezende2016,Lin2016}. Therefore, to answer the fundamental question of whether the spin angular momentum is transferred by coherent magnons \cite{Khymyn2016,Tatara2019} or via diffusion of thermal magnons \cite{Rezende2016,Lin2016}, the coherence of the spin current must be verified. Measurements of \textit{ac} spin current at nanometre length scales would provide crucial insight, but until now have not been reported.

\begin{figure*} 
\centering
\includegraphics[width=0.9\columnwidth]{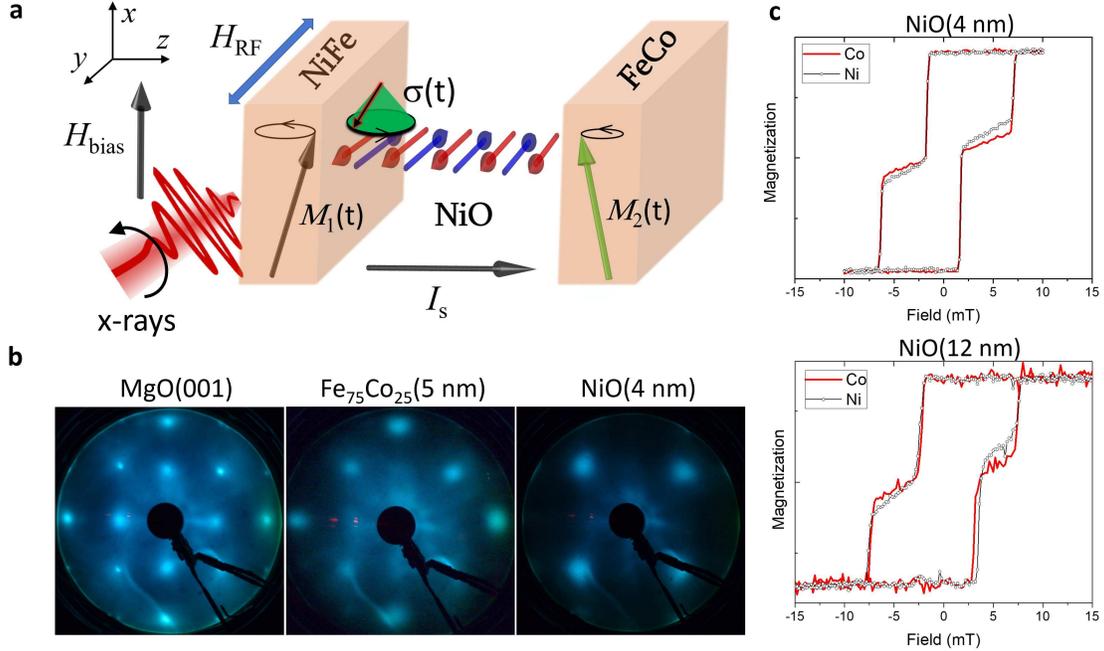}
   \caption{\textbf{Schematic diagram of the spin current transport and properties of NiO films.} \textbf{a}, Schematic diagram of the XFMR experiment and key concepts of spin pumping in a FM/AFM/FM multilayer structure. Precession of the magnetization $M_1$ of the NiFe source layer about the bias field $H_{\textrm{bias}}$ is induced by an in-plane rf magnetic field $H_{\textrm{RF}}$. The precession generates a spin current $I_{\textrm{S}}$ with time-dependent spin polarization $\sigma(t)$ that has both \textit{dc} (parallel to the x axis) and \textit{ac} (confined within the $\textit{yz}$ plane) components. The spin current propagates through the 90$^\circ$ coupled NiO layer and is absorbed at the NiO/FeCo interface, inducing precession of the FeCo magnetization $M_2$. By tuning the x-ray energy to the $L_3$ absorption edges of Ni and Co, the amplitude and phase of the precession, with respect to $H_{\textrm{RF}}$, can be detected independently for the NiFe and FeCo layers. \textbf{b}, LEED images acquired from the MgO(001) substrate, 5\,nm of $\text{Fe}_{\text{75}}\text{Co}_{\text{25}}$(001) and 4\,nm of NiO(001). \textbf{c}, Element specific XMCD hysteresis loops for Co and Ni acquired with the magnetic field perpendicular to the field cooling direction (i.e., perpendicular to FeCo[100]\,$||$\,NiO[110]) for NiFe/Fe/NiO(4\,nm)/FeCo and NiFe/Fe/NiO(12\,nm)/FeCo} samples.
  \label{fig1}
\end{figure*}

In this article, we present experimental detection of coherent spin current propagation through epitaxial NiO(001) layers. By employing element specific x-ray ferromagnetic resonance (XFMR) \cite{Marcham2013,Li2016,Laan2017}, the phase and amplitude of the magnetization precession within adjoining source and  sink ferromagnetic (FM) layers are detected, so the injection and transmission of pure \textit{ac} spin current through NiO can be inferred. Two different scenarios are explored: i) with NiO directly coupled to the FM layers so that the propagation of spin current through both NiO interfaces is assisted by interfacial exchange coupling; and ii) with the NiO decoupled from the sink layer by insertion of an additional non-magnetic (NM) spacer layer so that propagation of the spin current through the NM is detected via the spin-transfer torque (STT) acting at the NM/FM interface.

By tuning the x-ray energy to the absorption edge of the element of interest, the XFMR signals from the NiFe source layer (Ni $L_3$ edge) and the FeCo sink layer (Co $L_3$ edge) were measured separately, allowing direct measurement of the spin dynamics in each layer. Figure~\ref{fig1}a schematically shows the sample stack and the geometry of the spin transport within the NiO film. While previous studies concerning spin current propagation through NiO have focused on the time averaged \textit{dc} component, here the \textit{ac} component of the spin current \cite{Jiao2013,Li2016,Wei2014} is probed by measuring the magnetization precession of the FeCo sink layer \cite{Li2016}. Representative low energy electron diffraction (LEED) pattern images are shown in Fig.~\ref{fig1}b confirming the epitaxial growth of the FeCo and NiO layers, with the crystal orientation relation MgO[100]$||$FeCo[110]$||$NiO[100]. X-ray magnetic circular dichroism (XMCD) hysteresis loops for Co and Ni for samples with NiO thickness $d$ = 4 and 12\,nm, with the magnetic field applied in the sample plane perpendicular to the field cooling direction, are shown in Fig.~\ref{fig1}c. The split hysteresis loops result from the uniaxial anisotropy induced by field cooling. Both ferromagnetic layers switch at the same field values and are always collinear, as a result of the 90$^\circ$ coupling at the NiO interfaces (see Methods and Supplementary Information). No exchange bias is observed, as expected for an AFM with weak magnetic anisotropy and compensated spins.

\begin{figure*}
\centering
\includegraphics[width=1.0\columnwidth]{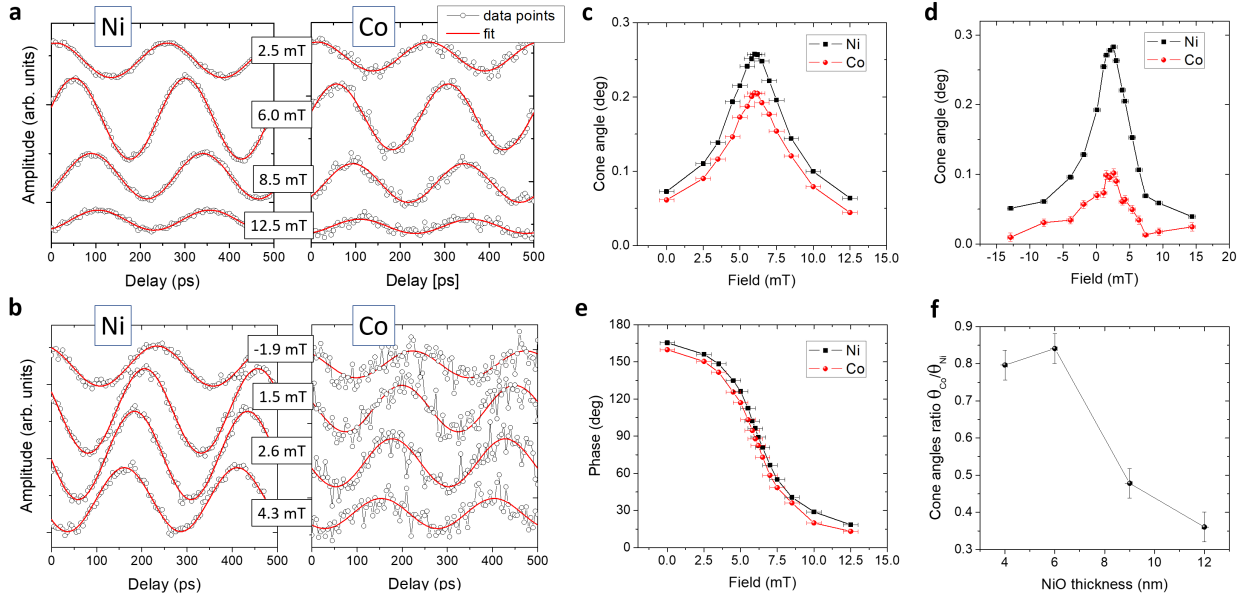} 
   \caption{\textbf{Detection of \textit{ac} spin current in directly coupled layers.} XFMR delay scans, with fitted sinusoidal functions, that reveal the precession of the Ni (NiFe layer) and Co (FeCo layer) moments at different bias fields for (\textbf{a}) NiFe/Fe/NiO(4\,nm)/FeCo and (\textbf{b}) NiFe/Fe/NiO(12\,nm)/FeCo. Amplitude of precession of the Ni and Co moments for (\textbf{c}) NiO(4\,nm) and (\textbf{d}) NiO(12\,nm). For all  samples with directly coupled layers, the FeCo and NiFe layers precess with the same phase, as shown in (\textbf{e}) for the sample with 4\,nm thick NiO. \textbf{f}, Ratio of the Co and Ni cone angles for different thicknesses of NiO layer.}
  \label{fig2}
\end{figure*}

Precession of the NiFe and FeCo moments was measured in directly coupled  NiFe/Fe/NiO($d$)/FeCo structures in the vicinity of the NiFe resonance at 4\,GHz excitation frequency for different NiO thicknesses. Both the amplitude and phase of precession were extracted by fitting the XFMR delay scans to a sine wave, as shown for $d$ = 4 and 12\,nm samples in Fig.~\ref{fig2}a and Fig.~\ref{fig2}b, respectively. To quantify the amplitude of the precession at different fields the FMR precession cone angle was estimated by taking the ratio between the XFMR dynamic signal ($I_{\text{XFMR}}$) and the XMCD static signal ($I_{\text{XMCD}}$) as follows $\theta$ = arctan($I_{\text{XFMR}}$/$I_{\text{XMCD}}$) \cite{Li2016,Laan2017}. For $d=$ 4 and 12\,nm (Fig.~\ref{fig2}c,d), both layers exhibit a FMR peak with the resonance field $\mu_0 H_r = (6.1 \pm 0.5)$\,mT and $\mu_0 H_r = (2.1 \pm 0.5)$\,mT, respectively. The FeCo FMR frequency/field mode is well separated from that of NiFe (as verified by vector network analyzer (VNA)-FMR) and therefore the observed precession of the FeCo must be induced by the NiFe precession. Notably, both layers precess exactly in-phase while the phase changes by 180$^\circ$ as the field is swept through the resonance (Fig.~\ref{fig2}e). The in-phase precession was observed for all NiO thicknesses studied. The precession induced in the FeCo layer results from \textit{ac} spin current being pumped by the NiFe precession and propagating through the NiO layer. In contrast to the NM/AFM/NM and FM/AFM/NM structures in which propagation of spin current has previously been studied, here AFM moments are coupled at both interfaces and interfacial exchange coupling  contributes to the transfer of spin angular momentum in each case. The ratio of the cone angles $\theta_{\text{Co}}/\theta_{\text{Ni}}$ of the sink and the source layer precession at $H_r$ can be used as a measure of the \textit{ac} spin current $I_{\text{AC}}$ propagating through the NiO. The ratio reaches a maximum at $d$ = 6\,nm and decays exponentially for larger $d$ values (Fig.~\ref{fig2}f), in a similar manner as in previous studies of \textit{dc} spin current \cite{Wang2014,Hahn2014,Lin2016}. This non-monotonic thickness dependence suggests that spin current plays a significant role, and argues against the action of a single precessional mode of the exchange coupled layers, for which a monotonic thickness dependence would be expected.

\begin{figure*}
\centering
\includegraphics[width=0.7\columnwidth]{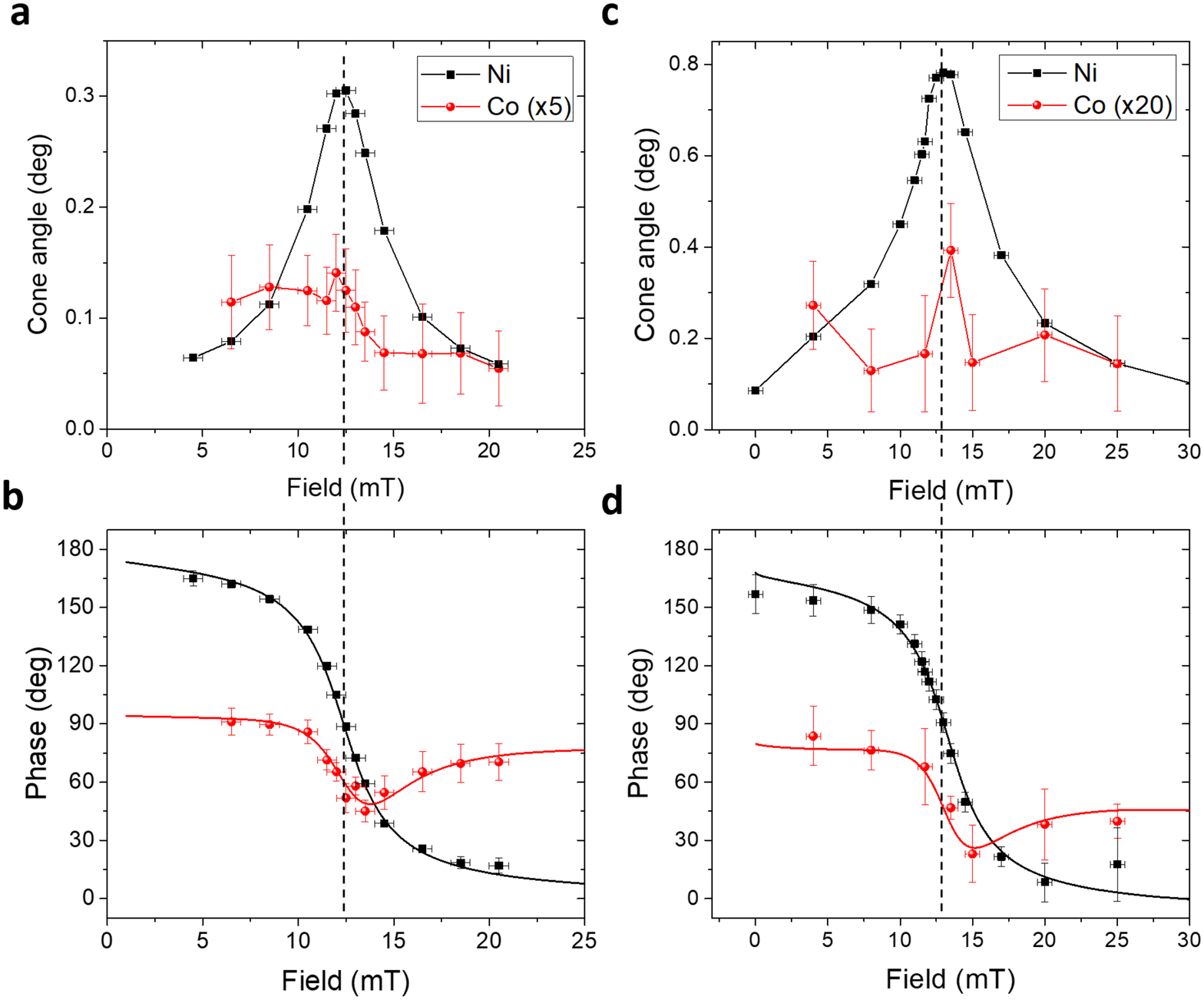} 
   \caption{\textbf{Detection of \textit{ac} spin current for samples with NM spacer layers.} \textbf{a},\textbf{c}, Amplitude and (\textbf{b},\textbf{d}) phase of the magnetization precession for (\textbf{a},\textbf{b}) NiFe/Fe/NiO(2\,nm)/\textbf{Ag(5\,nm)}/FeCo and (\textbf{c},\textbf{d}) NiFe/Fe/NiO(4\,nm)/\textbf{Pd(5\,nm)}/FeCo. Dashed lines indicate the bias fields $\mu_0 H_r$\,=\,12.5 and 13.25\,mT for the sample with Ag and Pd, respectively. The phase variations in both layers show the bipolar change at the $H_r$ and were fitted using equation (1).}
  \label{fig3}
\end{figure*}

To further prove the existence of \textit{ac} spin current propagation through NiO layers, additional experiments were performed on samples with a metallic 5\,nm thick NM layer of Ag or Pd inserted between the FeCo and NiO in order to remove the interfacial exchange coupling between these layers. In Fig.~\ref{fig3}, amplitudes (a,c) and phases (b,d) of the magnetization precession of FeCo and NiFe are shown for NiFe/Fe/NiO(2\,nm)/Ag(5\,nm)/FeCo (Fig.~\ref{fig3}a,b) and NiFe/Fe/NiO(4\,nm)/Pd(5\,nm)/FeCo (Fig.~\ref{fig3}c,d). While the behaviour of the NiFe is very similar as in the case of coupled layers shown in Fig.~\ref{fig2}, both the amplitude and phase of the magnetization precession of the FeCo are very different upon insertion of the NM layer. The amplitudes are substantially reduced as the spin current must pass through an additional interface and 5\,nm of NM material. In particular, the spin diffusion length in Pd is of order 2 - 10\,nm \cite{TaoFengMiaoEtAl2014} and comparable to the Pd layer thickness, while it is typically hundreds of nm in Ag \cite{Kimura2007}. Also, the mechanism of transfer of spin angular momentum to the FM is different, and is assumed to result mainly from spin transfer torque (STT) \cite{Marcham2013,Li2016} rather than interfacial exchange coupling. Most importantly, the phase of the FeCo precession undergoes a bipolar phase variation as the field is swept through the NiFe resonance (Fig.~\ref{fig3}b,d). This bipolar behaviour results from the FeCo layer precession being driven by the total torque due to the rf-field plus the \textit{ac} spin current \cite{Marcham2013,Li2016}. Although the bipolar phase variation can be clearly observed for all the measured samples with NM spacer layers, one cannot neglect interlayer exchange or dipolar coupling, which is still present in samples without and with 2\,nm thick NiO, as can be deduced from magnetometry and XMCD hysteresis loops (see Fig.~S3 in Supplementary Information). Regardless of the origin of the coupling (exchange or dipolar), it leads to a unipolar rather than a bipolar variation of the phase \cite{Marcham2013,Li2016,Baker2016}. To obtain a more quantitative measure of the different contributions to the FeCo layer precession, the XFMR results can be modeled by a linearized macrospin solution of the Landau-Lifshitz-Gilbert equation that incorporates both interlayer coupling and the STT due to spin pumping \cite{Marcham2013,Heinrich2003,Li2016}. The relative phase variation of the FeCo layer can be expressed by \cite{Li2019}: 
\begin{eqnarray}\label{phase}
&& \tan(\phi_{\text{FeCo}} - \phi^{0}_{\text{FeCo}}) = \nonumber \\
&& = \frac{\beta_{\text{cp}}\sin^2\phi_{\text{NiFe}}-\beta_{\text{sc}}\sin\phi_{\text{NiFe}}\cos\phi_{\text{NiFe}}}{1+\beta_{\text{cp}}\sin\phi_{\text{NiFe}}\cos\phi_{\text{NiFe}} + \beta_{\text{sc}}\sin^2\phi_{\text{NiFe}}},
\end{eqnarray}
where $\phi^{0}_{\text{FeCo}}$ corresponds to the phase of the FeCo precession driven by the rf-field alone, $\phi_{\text{NiFe}}$ is the phase of the NiFe precession and $\beta_{\text{cp}}$ and $\beta_{\text{sc}}$ are dimensionless parameters expressing the contributions of the interlayer coupling and the spin current, respectively \cite{Li2019}. The phase data were fitted with $\beta_{\text{cp}}$ and $\beta_{\text{sc}}$ as fitting parameters, with damping constants $\alpha_{\text{NiFe}}$ = 0.003, $\alpha_{\text{FeCo}}$ = 0.01 ($\alpha_{\text{FeCo}}$ = 0.02 for the sample with Pd spacer) and saturation magnetizations $\mu_0 M_{\text{s}}$ = 0.8\,T and 2.4\,T for NiFe and FeCo, respectively, which fall within the range of values reported in the literature \cite{Lee2017,Glowinski2019,Li2016}. The experimental data are well reproduced for samples with different thicknesses of NiO (see Fig.~\ref{fig3}b,d and section S3 of the Supplementary Information).

\begin{figure*}
\centering
\includegraphics[width=0.7\columnwidth]{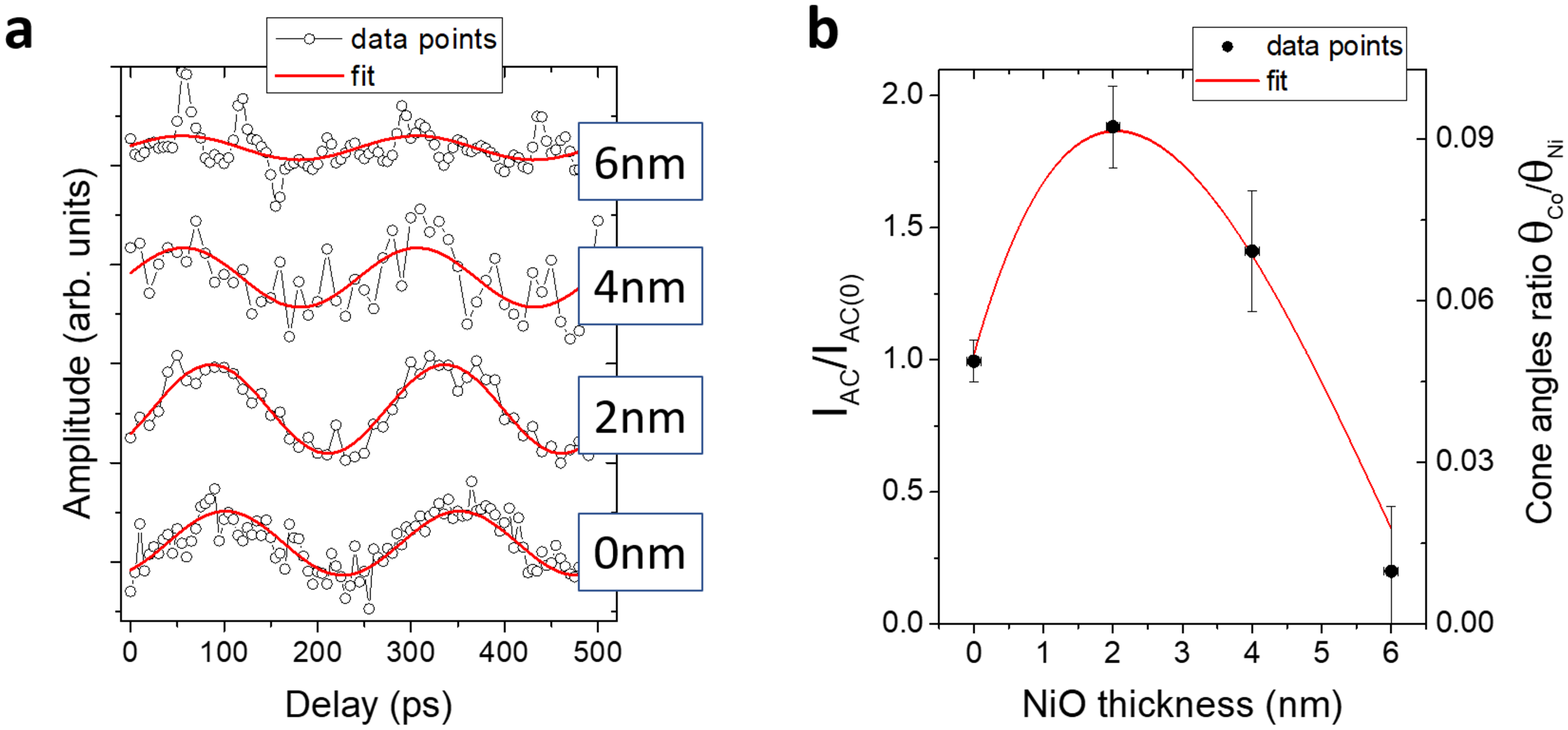} 
   \caption{\textbf{Dependence of the spin current on NiO thickness.} \textbf{a}, XFMR delay scans for the FeCo layer at $H_r$ for samples with a 5\,nm thick Ag spacer layer and NiO thickness $d$ = 0, 2, 4, 6\,nm. \textbf{b}, Dependence on $d$ of the spin current transmission efficiency $I_{\text{AC}}/I_{\text{AC(0)}}$ (the $\theta_{\text{Co}}/\theta_{\text{Ni}}$ cone angle ratio). The solid line shows a fit to the model of Khymyn \textit{et al.} \cite{Khymyn2016}.}
  \label{fig4}
\end{figure*}

The dependence of the \textit{ac} spin current on the NiO film thickness was explored next. In Fig.~\ref{fig4}a, XFMR delay scans for FeCo/Ag(5nm)/NiO($d$)/Fe/NiFe samples with different thickness $d$ are shown at their resonant fields (maximum amplitudes). An oscillatory XFMR signal associated with the precession can be clearly observed for all the investigated samples and appears to be largest for $d$ = 2\,nm, suggesting an enhancement in comparison to the case without NiO. The spin transfer efficiency can be quantified via the ratio of cone angles, $I_{\text{AC}}=\theta_{\text{Co}}/\theta_{\text{Ni}}$,  as in the case of directly coupled layers. In order to simplify comparison with previous experiments and theoretical models concerning \textit{dc} spin current \cite{Wang2014,Rezende2016}, values of $I_{\text{AC}}$ are normalized to $I_{\text{AC(0)}}$, the value for the sample without NiO. The dependence of $I_{\text{AC}}/I_{\text{AC(0)}}$ on NiO thickness for samples with 5\,nm Ag spacer layer is shown in Fig.~\ref{fig4}b. It is observed that the \textit{ac} spin current nearly doubles upon insertion of 2\,nm of NiO and then decreases for larger thickness values. This result closely resembles the enhancement of the \textit{dc} spin current measured by inverse spin Hall effect (ISHE) experiments \cite{Wang2014,Hahn2014,Lin2016}. 

The presented results confirm the propagation of \textit{ac} spin current through an AFM layer at a temperature well below the N\'eel temperature $T_N$ = 520\,K (for bulk NiO), suggesting that the spin current is mediated by coherent excitations of the AFM that have GHz frequency.  Despite the frequency mismatch between the GHz spin current and the THz magnon spectrum of the AFM, spin angular momentum may be transported by means of GHz evanescent AFM spin waves \cite{Khymyn2016} or AFM magnon pair propagation \cite{Tatara2019}.  Before examining one of these models in detail it is useful to first consider conceptually why such modes might be expected.  For a FM/AFM/FM system with interfacial exchange coupling it seems obvious that, if the exchange interaction throughout the structure is very strong, all the magnetic moments through the structure should be locked together, causing the magnetizations of the two FM layers to rotate in unison when an external magnetic field is applied.  However, if the exchange interaction within the AFM layer is reduced, the AFM magnetic order will no longer be rigid and there should be a small but finite twist in the alignment of the magnetic moments through the thickness of the AFM layer as the magnetization of the source FM layer is rotated.  When the precessional character of the motion is also taken into account, this twist is in fact synonymous with the evanescent AFM spin wave modes proposed by Khymyn \emph{et al.} \cite{Khymyn2016}.  The amplitude of the sink layer response has been shown to have a similar non-monotonic dependence upon the NiO thickness both with and without the insertion of a NM spacer into the stack.  This demonstrates that GHz evanescent AFM spin waves mediate the spin current in both cases.

Following the model of Khymyn \emph{et al.} \cite{Khymyn2016} let us now consider the spin current propagated through bulk-like NiO by a pair of evanescent AFM spin waves with eigenfrequencies of $\omega_1$ = 240\,GHz and $\omega_2$ = 1.1\,THz that correspond to easy plane and out of plane magnon modes, respectively. Based on the observed static 90$^\circ$ interfacial coupling and x-ray magnetic linear dichroism (XMLD) measurements of the magnetic order of the NiO layers, it is assumed that (001) is the easy plane and [001] is the hard axis, so that two modes with frequencies $\omega_1$ and $\omega_2$ may be excited by coupling to the NiFe source layer. For the excitation frequency of 4\,GHz used in the experiment, $\omega_1$ and $\omega_2$ correspond to two evanescent linearly polarized spin wave modes with penetration lengths of 22\,nm and 5\,nm, respectively. Using equation (9) in \cite{Khymyn2016}, the best fit is obtained when the phase shift between the two spin waves $\psi$\,=\,$\pi/2$\,$-$\,0.1\, (Fig.~\ref{fig4}b). The initial phase shift $\psi$ at the NiFe/NiO interface is expected to be exactly $\pi/2$ in the present case since the precessing source layer has \textit{ac} components within the $\textit{yz}$ plane (see Fig.~\ref{fig1}a). While it is well established that the injected spin current can induce oscillations of the NiO spins in the \textit{z} direction in the present geometry, the 90$^\circ$ exchange coupling allows oscillations to be induced in the \textit{x} direction. As a consequence, the two linear evanescent modes with polarization along the $\textit{x}$ and $\textit{z}$ axes, with frequencies $\omega_1$ and $\omega_2$ respectively, are excited with a $\pi/2$ phase difference (Fig.~\ref{fig1}a). Therefore, the fit implies that these two evanescent wave modes differ in phase by about $-$0.1\,rad after propagating through 6\,nm of NiO. Note that in the experiment the spin current propagates through another 5\,nm of Ag spacer before it is detected in the sink layer, while the model describes the spatial distribution of the spin current inside the AFM layer. Furthermore, the AFM magnon frequencies of bulk NiO assumed in the model may well be different to those of the thin NiO(001) layers studied here, which are strained and possess modified magnetic anisotropy.

The evanescent spin wave model does not explicitly take temperature dependence into account, in contrast to the thermal magnon model \cite{Rezende2016}, which can naturally explain the enhancement of the spin current near the N\'eel temperature $T_N$ \cite{Qiu2016,Lin2016,Chen2016,Hou2017}. It must be noted however that the AFM magnon frequencies decrease as $T_N$ is approached \cite{Sievers1963,Moriyama2019}, which should result in an increase in the penetration of the evanescent spin waves and hence an increase of the spin current. Since  $T_N$ also decreases as the NiO thickness $d$ decreases, for small $d$ values, such as $d$ = 2\,nm where $T_N$ $\approx$ 375\,K \cite{Boeglin2009}, the evanescent spin waves are expected to mediate the spin current more effectively at RT. The close correspondence between the thickness dependence of the \textit{ac} spin current measured here and that of the \textit{dc} spin current reported previously suggests that both quantities are driven by the same mechanism - the excitation of evanescent AFM spin waves. Otherwise, if the enhancement of the \textit{dc} spin current was dominated by incoherent thermal magnons, its thickness dependence would be expected to be different to that of the \textit{ac} spin current. As the temperature approaches $T_N$, the population of thermal magnons increases significantly, and additional enhancement of the \textit{dc} spin current is expected. Further studies of \textit{ac} spin current in NiO in the vicinity of $T_N$ are desirable, but the construction of the XFMR apparatus does not currently permit the sample to be heated above RT. 

Other models have been proposed in which GHz spin current is mediated by THz frequency magnons within an adiabatic approximation \cite{Takei2015}. The present observation of \textit{ac} spin current propagation places an upper limit of about 100\,ps on the time required for a suitable THz magnon population to form.  Specifically, it is necessary to explain both how THz magnons can be excited in an AFM by the GHz precession of the magnetization in an adjacent exchange coupled FM, and also how such magnons interact within the AFM to form a well defined statistical distribution on the timescales required.  Finally, the measurements in this study were performed at room temperature, well below $T_N$ for the NiO layers studied, and it therefore seems unlikely that thermal THz magnons could lead to the significant excitation of the sink layer magnetization observed in the XFMR experiments. This was further verified by XFMR measurements  at lower temperatures for the NiFe/Fe/NiO(4nm)/FeCo sample, where no significant change in amplitude of the FeCo magnetization precession was observed down to 100\,K.

In summary, it has been demonstrated that \textit{ac} spin current of GHz frequency can efficiently propagate through epitaxial NiO layers of different thickness at room temperature. The \textit{ac} spin current is enhanced for NiO thicknesses less than 6\,nm, both with and without a non-magnetic spacer layer inserted into the stack, in a manner consistent with previously reported experimental measurements of \textit{dc} spin current and theoretical studies. The results show that the propagation of spin current through NiO layers is mediated by evanescent antiferromagnetic spin wave modes of GHz frequency rather than THz frequency magnons.  The coherence of the spin current paves the way to applications in both spintronics and magnonics in which the phase of a spin wave or current is a well defined variable.  Confirmation of the evanescent spin wave mechanism also opens the door to the construction of spin current amplifiers that exploit the reservoir of angular momentum that resides within the lattice.

\begin{flushleft}
\textbf{Methods}
\end{flushleft}
\textbf{Sample fabrication.} The present study focuses on epitaxially grown NiO(001) embedded in the structure $\text{Ni}_{\text{80}}\text{Fe}_{\text{20}}$(25)/Fe(1)/NiO($d$)/$\text{Fe}_{\text{75}}\text{Co}_{\text{25}}$(5) (throughout the manuscript these stoichiometries are abbreviated by NiFe and FeCo, respectively) with different NiO thickness $d$ = 4, 5, 6, 9 and 12\,nm, grown on MgO(5)/MgO(001) and covered with a MgO(3) capping layer (thicknesses in nm). The 1\,nm thick Fe layer was grown on top of the NiO layer to form an epitaxial layer and provide a smooth interface. Another series of samples with an additional 5\,nm thick non-magnetic layer of Ag or Pd inserted between the NiO(\textit{t}) and $\text{Fe}_{\text{75}}\text{Co}_{\text{25}}$(5) layers was grown to fully suppress the interfacial exchange coupling between these layers. The quality and crystalline order of particular layers was confirmed by low energy electron diffraction (LEED) (Fig.~\ref{fig1}b). The majority of the samples were field-cooled from 550\,K to RT in a 1\,T field applied along the FeCo[100]\,$||$\,NiO[110] axis. For one set of samples, a 60\,mT bias field was applied during the growth of the NiO layers. No differences in the static and dynamic properties were observed between these samples and the field-cooled samples.

\medskip
\noindent
\textbf{Interfacial exchange coupling and static magnetic properties.} Based on x-ray magnetic linear dichroism (XMLD) measurements in which the x-ray polarization direction was varied, as well as x-ray magnetic circular dichroism (XMCD) hysteresis loops acquired in a transverse geometry, i.e., with the x-ray wavevector perpendicular to the applied field,  it was found that the NiO moments lie in-plane and are coupled at 90$^\circ$  to the moments in the adjacent FeCo and NiFe ferromagnetic layers, as shown schematically in Fig.~\ref{fig1}a. Perpendicular coupling of the FM and AFM moments is energetically most favourable for the fully compensated interface \cite{Koon1997,Cheng2018a,Laan2011} and attests to the high quality of the NiO layers while also excluding direct exchange coupling due to the formation of pinholes (see Supplementary Information for more details regarding 90$^\circ$ coupling).  Prior to the synchrotron experiments, the samples were characterized by vibrating sample magnetrometer (VSM) and vector network analyzer ferromagnetic resonance (VNA-FMR) measurements. The hysteresis loops reveal a 4-fold anisotropy with an additional small uniaxial anisotropy with easy axis along the field cooling direction (i.e., along FeCo[100]\,$||$\,NiO[110]). As a consequence, square or split hysteresis loops (Fig.~\ref{fig1}c) are measured with the magnetic field applied parallel and perpendicular to FeCo[100], respectively.   

\medskip
\noindent
\textbf{Time-resolved measurements.} X-ray ferromagnetic resonance (XFMR) experiments \cite{Marcham2013,Li2016,Arena2006,Laan2017} were carried out on beamlines 4.0.2 of the Advanced Light Source (USA) and I10 at the Diamond Light Source (UK), by monitoring the time delay between a synchronized radio frequency (rf) magnetic field (pumping the spin precession) and circularly polarized x-ray pulses (probing the oscillatory magnetization component along the x-ray wavevector). The x-ray incidence angle was set to 50$^\circ$ with respect to the sample normal. The sample was placed face down on a coplanar waveguide (CPW) with a countersunk hole of 500\,$\mu$m diameter allowing the incident x-ray beam to access the surface of the sample,  while the transmitted x-rays were converted through x-ray excited optical luminescence in the MgO substrate, with the emitted light detected by a photodiode mounted behind the sample. Further details of the XFMR experiment can be found in reference \cite{Laan2017}. An excitation frequency of 4\,GHz was used for all the measurements presented here. 

\medskip
\noindent
\textbf{Acknowledgments:} The authors acknowledge the Engineering and Physical Sciences Research Council (EPSRC) under Grant
Numbers EP/P021190/1, EP/P020151/1, and EP/P02047X/1. This work was supported by US Department of Energy, Office of Science, Office of Basic Energy Sciences, Materials Sciences and Engineering Division under Contract No. DE-AC02-05-CH11231 (van der Waals heterostructures program, KCWF16). Beamtime awarded on I10 at the Diamond Light Source (SI17745-1, SI19116-1 and SI20760-1) is acknowledged. This research used resources of the Advanced Light Source, which is a DOE Office of Science User Facility under contract no. DE-AC02-05CH11231. T.N. acknowledges JSPS Overseas Research Fellowships. D.G.N. acknowledges support via the EPSRC Centre for Doctoral Training in Metamaterials (Grant No. EP/L015331/1).

\medskip
\noindent
\textbf{Author contributions}
M.D., T.N., D.M.B., A.F., D.G.N. and C. K. performed the measurements; Q.L., M.Y. and Z.Q.Q. fabricated the samples; M.D. and T.N. analysed the data; M.D. and R.J.H. wrote the manuscript; all authors contributed to the discussion and approved the manuscript; R.J.H. and G.v.d.L. supervised the project. 

\medskip
\noindent
\textbf{Additional information}
Correspondence and requests for materials should be addressed to M.D. or R.J.H.

\medskip
\noindent
\textbf{Competing financial interests}
The authors declare no competing financial interests.

\bibliographystyle{naturemag}

\end{document}